\documentclass[sigconf]{acmart}
\usepackage{amsmath}
\usepackage{graphicx}
\usepackage{multirow}
\usepackage{subfigure} 
\AtBeginDocument{%
  \providecommand\BibTeX{{%
    \normalfont B\kern-0.5em{\scshape i\kern-0.25em b}\kern-0.8em\TeX}}}

\copyrightyear{2023} 
\acmYear{2023} 
\setcopyright{acmlicensed}\acmConference[ICMR '23]{International Conference on Multimedia Retrieval}{June 12--15, 2023}{Thessaloniki, Greece}
\acmBooktitle{International Conference on Multimedia Retrieval (ICMR '23), June 12--15, 2023, Thessaloniki, Greece}
\acmPrice{15.00}
\acmDOI{10.1145/3591106.3592216}
\acmISBN{979-8-4007-0178-8/23/06}
 
\begin{document}

\title{Joint Geometric-Semantic Driven Character Line Drawing Generation}

\author{Cheng-Yu Fang}
\authornote{Both authors contributed equally to this research.}
\email{fangchengyuswu@163.com}
\orcid{0000-0002-6522-3710}
\author{Xian-Feng Han}
\authornotemark[1]
\authornote{Corresponding authors}
\email{xianfenghan@swu.edu.cn}
\orcid{0000-0002-4869-4537}
\affiliation{%
  \institution{Southwest University}
  \streetaddress{No.2 Tiansheng Road}
  \city{Chongqing}
  \state{}
  \country{China}
  \postcode{400700}
}

\renewcommand{\shortauthors}{}

\begin{abstract}
Character line drawing synthesis can be formulated as a special case of image-to-image translation problem that automatically manipulates the photo-to-line drawing style transformation. In this paper, we present the first generative adversarial network based end-to-end trainable translation architecture, dubbed P2LDGAN, for automatic generation of high-quality character drawings from input photos/images. The core component of our approach is the joint geometric-semantic driven generator, which uses our well-designed cross-scale dense skip connections framework to embed learned geometric and semantic information for generating delicate line drawings. In order to support the evaluation of our model, we release a new dataset including 1,532 well-matched pairs of freehand character line drawings as well as corresponding character images/photos, where these line drawings with diverse styles are manually drawn by skilled artists. Extensive experiments on our introduced dataset demonstrate the superior performance of our proposed models against the state-of-the-art approaches in terms of quantitative, qualitative and human evaluations. Our code, models and dataset will be available at Github.

\end{abstract}



\begin{teaserfigure}
  \includegraphics[width=\textwidth]{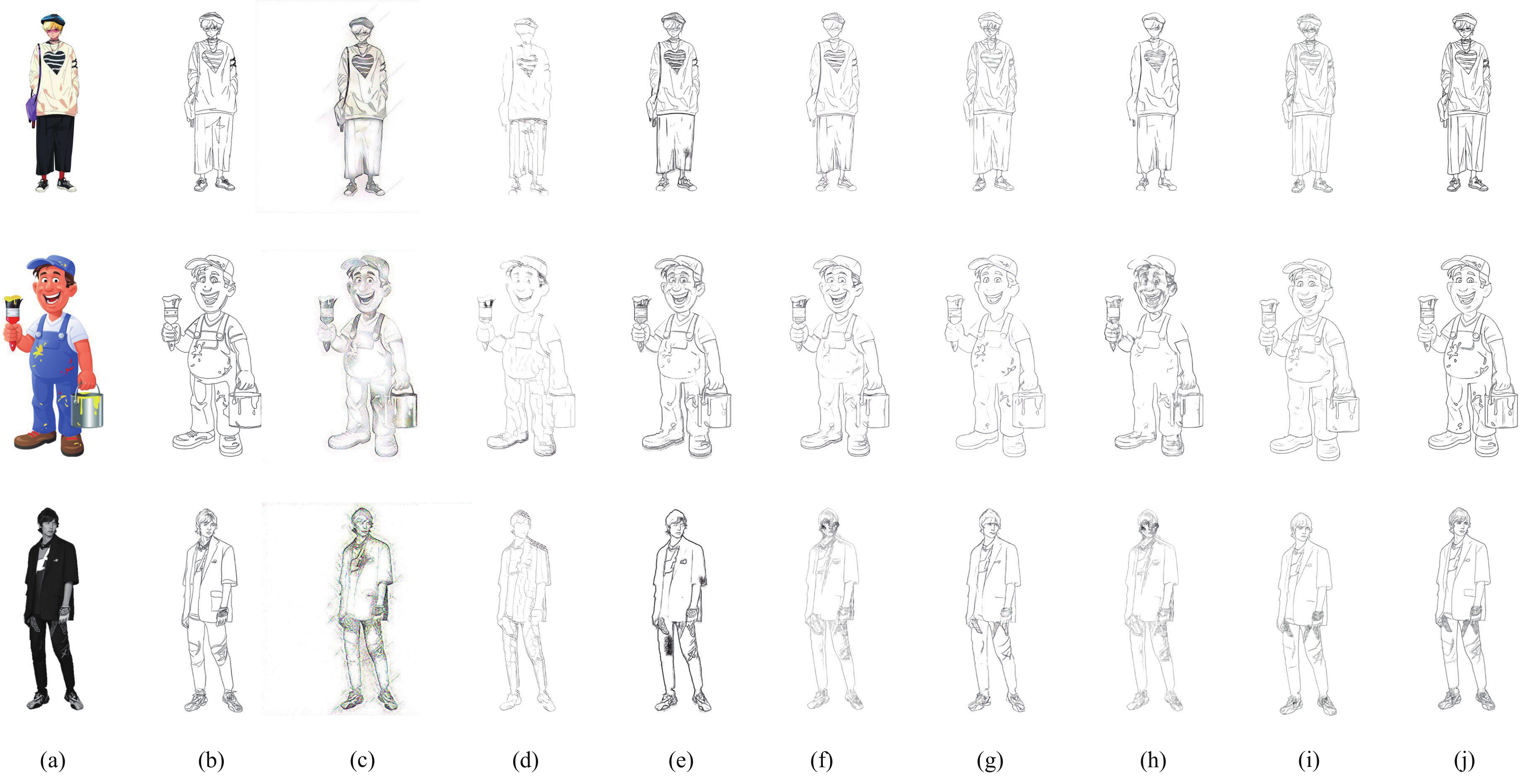}
  \caption{Qualitative comparison with state-of-the-art methods. All these models are trained on our built dataset. (a) Input photo/image; (b) Ground truth; (c) Gatys \cite{gatys2016image}; (d) CycleGAN \cite{zhu2017unpaired}; (e) DiscoGAN \cite{kim2017learning} ; (f) UNIT \cite{liu2017unsupervised}; (g) pix2pix \cite{isola2017image}; (h) MUNIT \cite{huang2018multimodal}; (i) Our baseline; (j) Our P2LDGAN.}
  \label{fig:topbanner}
\end{teaserfigure}

\maketitle

\section{Introduction}
Character Line Drawing, also named Line Art, aims to translate the information in photograph/image domain into a simplified representation domain with fundamental graphic components (e.g., lines or curves) to show the change of plane. It is an abstract and flexible form of art, whose applications include diverse scenes, such as entertainment \cite{yan2021isgan}, key art \cite{chen2020puppeteergan}, caricature \cite{jang2021stylecarigan} and computer generated animation \cite{mourot2021survey}. As we all know, high-quality line drawing generation typically requires professional artists or domain experts to spend considerable effort. And more details in structural lines mean more difficulties in drawing the line arts \cite{zheng2020learning}. Therefore, manually drawing character line arts is a labor-intensive, time-consuming and challenging task. And it is highly desired to development an automatic generation method that can assist artists or amateurs in drawing character line drawings.  

Recently, the great process of convolutional neural network (CNN), especially generative adversarial network (GAN), has popularized the image generation/translation tasks due to their powerful ability in synthesizing impressive, high-visual-quality, realistic and faithful images. Is it possible to automatically perform character photo-to-line drawing style translation with the help of these artificial intelligence techniques? In essence, this question can be considered as an image-to-image translation problem, which converts the character image/photo to line drawings representation domain. The GAN-based approaches are actually applicable to tackling this problem. 

However, to the best of our knowledge, there is no previous studies specifically developed for character line drawing creation. In addition, directly extending the state-of-the-art image translation methods (e.g., pix2pix \cite{isola2017image}, CycleGAN \cite{zhu2017unpaired}, \cite{liu2017unsupervised}) to this task cannot achieve high-quality character drawings due to the following issues. (1) Freehand character line drawings is abstract \cite{gao2020sketchycoco} and sparse \cite{yi2022quality}, totally different from other image translation applications. (2) Loss of semantic information results in unclear and imperfect lines.

Therefore, in this paper, we attempt to make these challenges resolvable. We introduce the first GAN-based end-to-end architecture, called P2LDGAN, aiming to automatically learn the cross-domain corresponding relations between images/photos and hand-drawn line drawings. The starting point of our P2LDGAN is an input real character photo, outputting a realistic hand-drawn character line drawing. To improve the generation quality with more details and clear lines, we design a joint geometric-semantic driven generator, in which the feature maps with different scales and information flows from encoding stage are densely concatenated into corresponding layers of decoder using cross skips to fuse geometric and semantic features for find-grained drawings generation. For discriminator, we adopt patch discriminator to improve the discriminative ability. 

In order to train and evaluate our proposed model, we introduce a new dataset, which consists of more than one thousand of character images/photos and line drawings pairs, where the line drawings are manually created by skilled artists we invite. We quantitatively and qualitatively compare our framework against the state-of-the-arts on this newly collected dataset, and experimental results show the superiority of proposed P2LDGAN. Finally, we perform the ablation studies to further validate the effectiveness of our key components.

Summarily, the main contributions of our study include,
\begin{itemize}
    \item To the best of our knowledge, this is probably the first work focusing on full body character line drawing generation study.
    
    \item We contribute a new photo-to-line drawing dataset including 1,532 pairs of character photos/images and corresponding well-aligned freehand line drawings provided by professional artists, to benefit the research on character line drawing creation.  
    
    \item We present the first joint geometric-semantic driven generative adversarial architecture with our well-designed cross-scale dense skip connections framework as generator for  automatic character line drawing generation in an end-to-end manner.
    
    \item Experimental results as well as perceptual study demonstrate the effectiveness of the proposed P2LDGAN. Compared to state-of-the-art methods, our model is more robust to different categories and styles of the input and can achieve significantly higher quantitative and qualitative performance. 
    
\end{itemize}

\begin{figure*}[htbp]
  \centering
  \includegraphics[width=\textwidth]{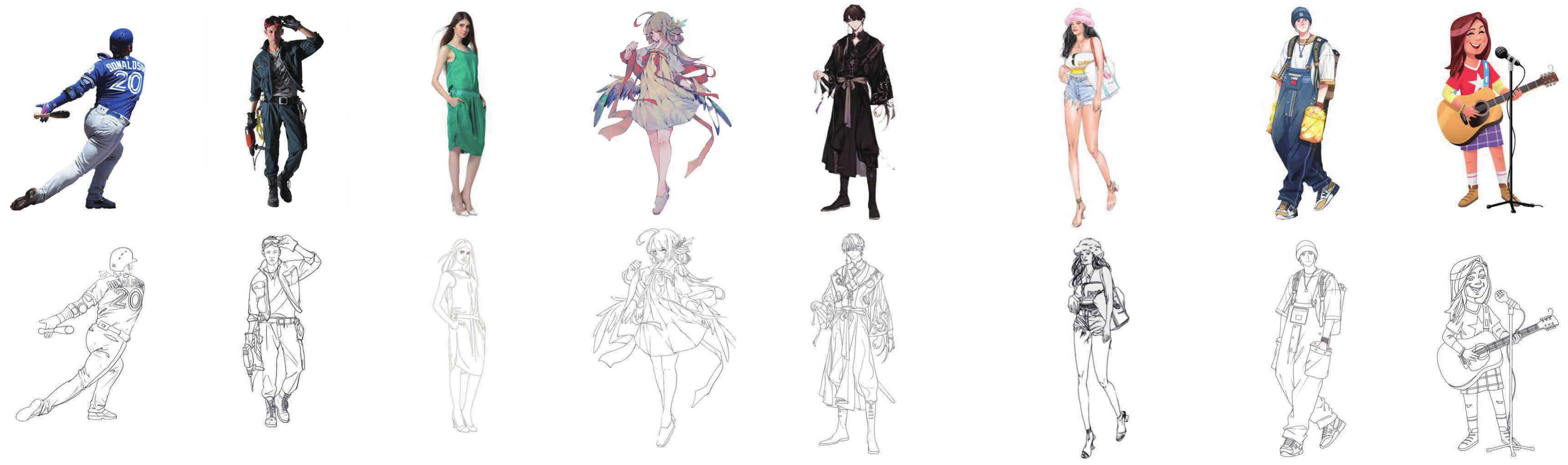}
  \caption{Representative image/photo-line drawing pairwise examples.  }
  \label{fig_example}
\end{figure*}

\section{Related Work}
\subsection{Image-to-Image Translation}
The image-to-image translation \cite{mao2022continuous}\cite{wang2022coarse}\cite{richardson2021encoding}\cite{pang2021image} can be formulated as an image generation function that maps the given source domain image into the desired target artistic style.  It has a wide range of applications, including image cartoonization \cite{dong2021cartoonlossgan}\cite{wang2020learning}, flat filling \cite{zhang2021user}\cite{wu2021towards}, face sketch synthesis \cite{li2021face}\cite{cao2022face} and image inpainting \cite{zhao2020uctgan}\cite{liu2021pd}. Particularly, the generative adversarial network based image translation has been receiving substantial interest. Based on conditional adversarial network, Pix2pix \cite{isola2017image} develops a general image translation framework for different tasks, like colorization,  sketch-to-portrait. However, the training instability makes it difficult to apply pix2pix to high-resolution image generation issue. Wang et al. \cite{wang2018high} introduced the multi-scale generator and discriminator models together with a new feature match loss to form pix2pixHD to achieve high-resolution results. Qi et al. \cite{qi2021face} developed a semantic-driven GAN with pix2pix as backbone, and designed an adaptive re-weighting loss to balance semantic classes' contributions. In contrast with learning from paired data, unsupervised image translation has recently attracted more and more attention. CycleGAN \cite{zhu2017unpaired} designs a cycle consistency constraint to conduct learning with unpaired inputs. Similarly, DualGAN  \cite{yi2017dualgan} also proposes a general-purpose image translation framework using a cyclic mapping. While DiscoGAN \cite{kim2017learning} couples two different GANs for bidirectional mapping between source and target domains. Based on the assumption that corresponding images from different domains share the same intermediate representation in a latent space, Liu et al. \cite{liu2017unsupervised} integrated the core ideas of GANs and variational autoencoders (VAEs) forming the UNIT framework to model each image domain. MUNIT \cite{huang2018multimodal} assumes that the representation of source and target images can be decomposed into a shared content space but different domain-specific style spaces. To achieve image translation, it performs a swapping combination of these content and style codes. 

\subsection{Deep Learning in Line Drawing}
The deep learning methodology has already pushed significant steps in a wide variety of visual processing tasks, which makes great contributions to line drawing related applications, including line drawing colorization \cite{yuan2021line}, artistic shadow creation \cite{zhang2021smartshadow}, manga structural line extraction \cite{li2017deep}, and face portrait line drawing \cite{yi2020unpaired}. Zhang et al. \cite{zhang2021user} proposed a split filling mechanism to perform line art colorization. They first spilt the input user scribbles into several groups for influence areas estimation, then a data-driven color generation process is conducted for each group. These outputs are finally merged to form the high-quality filling results. Zheng et al. \cite{zheng2020learning} first designed a ShapeNet to learn the 3D geometric information from line drawings, then a RenderNet is performed to produce 3D shadow. The SmartShadow \cite{zhang2021smartshadow} application consists of three data-driven tools, a shadow brush used to determine the areas the user want to create shadow, a shadow boundary brush to control the shadow boundaries, and a global shadow generator for the entire image shadow generation based on the estimated global shadow direction. Li et al. \cite{li2017deep} took advantage of the ideas of residual network and symmetric skipping network to build their CNN based method to solve problem of structural line extraction from manga image. To generate portrait drawings in multiple styles and unseen style, Yi et al. \cite{yi2022quality} designed a novel portrait drawing generation architecture using a asymmetric cycle structure, where a regression network is first adopted to calculate the quality score for an APDrawing. Based on this model, they defined a quality loss to guide the network to generate high-quality APDrawings. Im2Pencil \cite{li2019im2pencil} and \cite{zhang2021generating} treat the procedure of line drawing generation as an independent subtask to facilitate manga and Illustration synthesis, respectively.  

Different from these methods, we apply the generative adversarial network to a new problem, i.e. our P2LDGAN mainly concentrates on finding cross-domain relations between character images/photos and hand-drawn character line drawings using paired data.  

\section{Data Preparation}

Our photo/image-to-line drawing converter can be fundamentally posed as supervised image-to-image translation problem which heavily relies on significant training samples, therefore, it is necessary to construct a dataset to help our model learn an accurate mapping to bridge the real character photo/image and freehand character line drawing domains. 

Specifically, we first collect high-resolution character images/photos from the internet mainly covering five categories, namely male, female, manga/cartoon male, manga/cartoon female, and others, to enrich or diversify the data. Then, to meet the demands of different artistic styles, we invite experienced artists and many skilled students majored in visual communication design to manually draw the character line drawings at the same scale as the given source images/photos using specific application and digital devices. Finally, to standardize the gathered images/photos and hand-drawn line arts, we preprocess them to the size of 1024 $\times$ 1024 pixels, and fine tune the structural lines manually to form the strictly aligned image/photo-line drawing pair (describing the same character) with the help of professional artists and image processing software.

We carefully select 1,532 pairs of high-quality character line drawings paired with real images/photos to construct our final dataset. Table 1 illustrates the distributions of our collected dataset. Figure \ref{fig_example} shows representative character image/photo-line drawing pairs. In order to facilitate the learning and evaluation, we choose to split the dataset into two disjoint subsets, i.e. 70\% for training and the remaining 30\% as testing set. The data will be released to the public for research related to line drawing.

\small
\begin{table}[]
\centering
\caption{The Composition of our constructed benchmark dataset. The bottom row shows the number of image/photo-line drawing pairs contained in each category.}
\begin{tabular}{|c|c|c|c|c|c|}
\hline
 Category& Male & Female  & Cartoon Male & Cartoon Female   & Others  \\ \hline
 No. & 388 & 422 & 210 & 426 & 86 \\
 \hline
\end{tabular}
\label{tab: dataset}
\end{table}

\section{Character Photo-to-Line Drawing Translation Architecture}

\begin{figure*}[htbp]
  \centering
  \includegraphics[width=\textwidth]{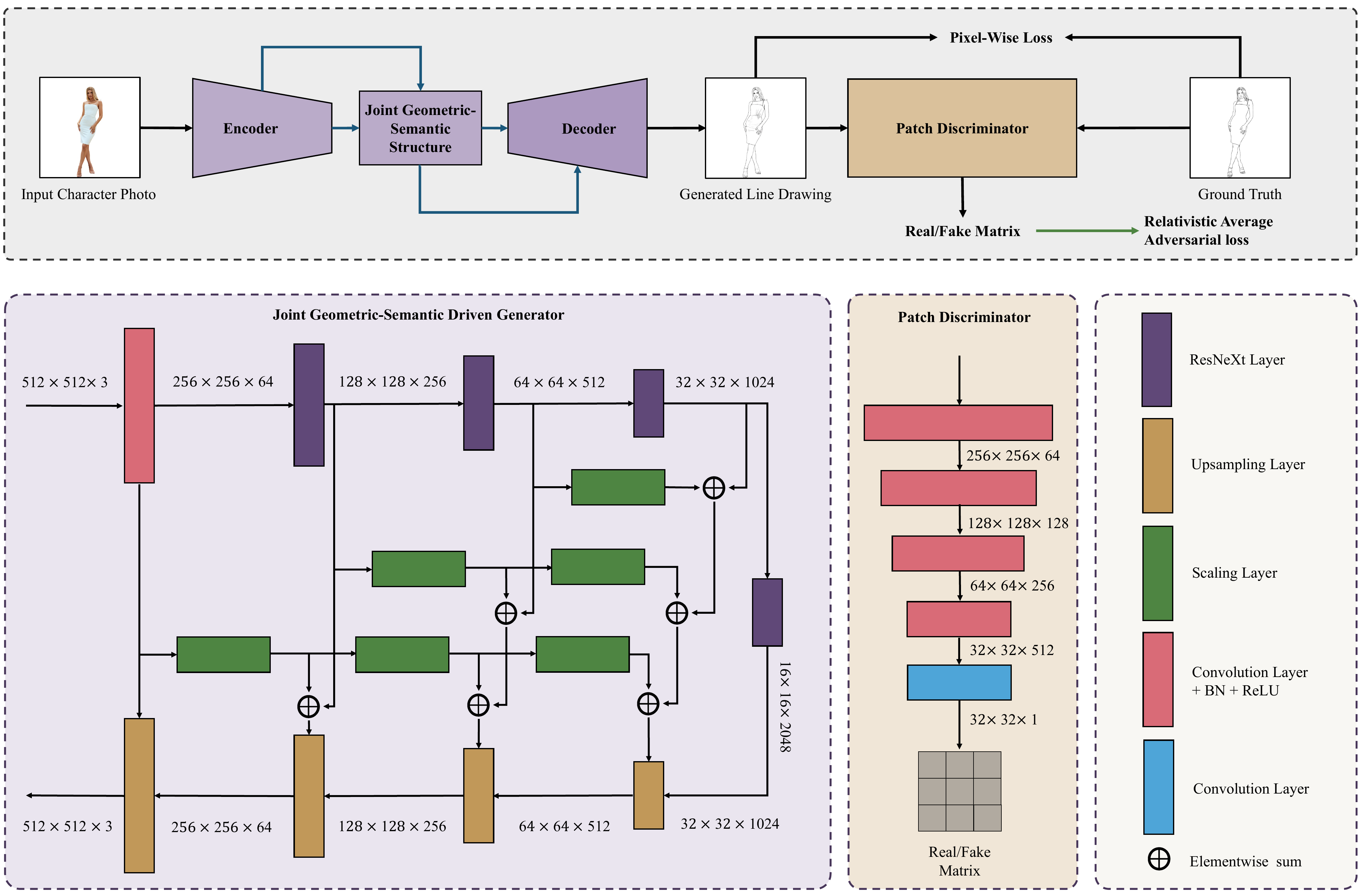}
  \caption{Overview of our high-quality photo-to-line drawing translation architecture (P2LDGAN). The bottom-left is the proposed multi-scale cross skip-connection module, the core component of our joint geometric-semantic driven generator. The bottom-middle displays the structure of patch discriminator.}
  \label{fig_framework}
\end{figure*}

\subsection{Overview}
We attempt to generalize the application of generative adversarial framework to the  photo-to-line drawing style conversion problem, i.e. automatic artistic character line drawings generation from real images/photos. Given the well-aligned source-reference pair $\{p_{i}, l{i}\}_{i=1}^{N}$, where $p_{i}$ and $l_{i}$ belong to character photo domain $\mathcal{P}$ and line drawing domain $\mathcal{I}$, respectively. $N$ denotes the number of image pairs. The goal of our model is to learn a mapping function $\mathcal{F}$ that automatically discover the domain relations between line drawing $\hat{l}_{i}$ and corresponding character photo. 

\begin{equation}
    \hat{l}_{i} = \mathcal{F}(p_{i}, l_{i})
\end{equation}

Figure \ref{fig_framework} illustrates the details of our developed P2LDGAN for this cross domain corresponding learning, which consists of (1) a joint geometric-semantic driven generator G, and (2) a character line drawing discriminator D. In the following, we give a detailed description of our method.
\subsection{Joint Geometric-Semantic Driven Generator}
The character line drawings describe the characters' clear and critical features using structural lines representation. Therefore, both geometric and semantic information play a fairly important role in synthesizing vital details in drawings. Previous state-of-the-art image translation models mainly adopt U-net framework with skip connections. However, these methods only combine the feature maps of the same scale from encoding and decoding stage, lacking geometric and semantic information fusion, which limits the generation quality.

Therefore, in order to progressively propagate the geometric and semantic information into the output line drawing, we development a simple but effective framework, named cross-scale dense skip connections module, which is the core of our joint geometric-semantic driven generator.

Our generator fundamentally is an encoder-decoder architecture, where the encoding stage compresses the rich information in character photo into latent representation, while the decoding network constructs the desired line drawings from encoded representation. As shown in the bottom-left part of Figure \ref{fig_framework}, we adopt the pre-trained ResNeXt-50 as encoder because of its simpleness, modularization, efficiency and higher learning capability. Specifically, we extract the stages conv2-conv5 from ResNeXt as our encoding layer (i.e. ResNeXt Layer), which can be formulated as,  

\begin{equation}
    \mathcal{F}_{i} = \mathcal{F}_{i-1} + \sum_{j=1}^{C}\mathcal{T}_{j}(\mathcal{F}_{i-1})
\end{equation}
Where $i$ indicates the $i$th layer of the encoder, $\mathcal{F}_{i-1}$ and $\mathcal{F}_{i}$ denoted the input and output. $C$ in the number of groups. $\mathcal{T}$ means transformation function, here we use a sequence of $1\times1$, $3\times3$ and $1\times1$ convolutions.

We input a character image/photo of size $512 \times 512$ into encoder, and extract feature maps from each ResNeXt layer together with our decoding layer to build the cross-scale skip connections model to fuse and propagate information of different levels of abstractness. To be specific, suppose our generator has $n$ layers in total, the input of the current decoding layer $n-i+1$ is the combination between the outputs of the previous layer and corresponding encoding layers having the same or larger resolutions. The formulation is defined as,

\begin{equation}
    \mathcal{I}_{n-i+1} = \mathcal{F}_{n-i} \oplus \mathcal{F}_{i} \oplus PDS(\mathcal{F}_{1}) \oplus \dots \oplus PDS(\mathcal{F}_{i-1}) 
\end{equation}

Where $\mathcal{F}_{n-i}$ represents the previous $n-i$th layer's output, $\mathcal{F}_{1}$ to $\mathcal{F}_{i}$ denote the outputs of the 1st to $i$th encoding layers with scales larger than or equal to that of the $n-i+1$ layer. PDS() refers to progressive scaling operation to scale feature maps using convolutions. $\oplus$ is the element wise sum operation. Subsequently, the fused multi-scale features flows through the decoding layer to perform geometric and semantic information propagation into a higher-resolution maps.   

\begin{equation}
 \mathcal{F}_{n-i+1} = UP(\mathcal{I}_{n-i+1})
\end{equation}
 Here, UP() operation is used to upsample the feature maps by a factor of 2 using nearest neighbor algorithm. Through our cross skip connection mechanism, feature maps from each encoding layer can be embedded into different decoding layers to strengthen semantic and geometric information integration and improve feature propagation across encoder and decoder, so that we can directly train our generator to learn character line drawing modality and translation the real character image/photo into line drawing with details preservation.

\subsection{Discriminator}
For discriminator network, its task is to discriminate the generated character line drawings from ground truth. Here, we adopt the PatchGAN exploited in \cite{isola2017image} to classify whether the $32 \times 32$ patches are real or not. The bottom-middle part of Figure \ref{fig_framework} displays discriminator architecture. We can observe that each discriminator block is composed of $4\times4$ Convolution, InstanceNorm and LeakyReLU with a slop of 0.2. Such a discriminator contributes to high-quality line drawings generation since it makes our P2LDGAN pay more attention to detailed content in patches, and can process images of any size with fewer parameters \cite{zhu2017unpaired}.

\subsection{Objective Functions}
\textbf{Adversarial Loss}. The generator $G$ creates character line drawing $\hat{l}$ from image $p$ in character photo domain $\mathcal{P}$, which can not be distinguished by discriminator $D$, while the goal of discriminator $D$ is to discriminate between translated drawings and real samples $l$ from $\mathcal{I}$. Motivated by the significant stability and high-quality generation capability of relativistic
average GANs \cite{jolicoeur2018relativistic}, we introduce the following Adversarial loss to supervise our P2LDGAN for more realistic line drawing synthesis.

\begin{equation}
    \begin{split}
            \mathcal{L}_{G}^{P2LDGAN} = \mathbb{E}_{p\sim \mathcal{P},l\sim \mathcal{I}}\left [ f_{1}(D(G(P))-f_{2}((D(l))), f_{3}(f_{4}(p))) \right ] 
    \end{split}
\end{equation}

\begin{equation}
    \begin{split}
            \mathcal{L}_{D}^{P2LDGAN} = \mathbb{E}_{p\sim \mathcal{P},l\sim \mathcal{I}}\left [ f_{1}(D(l)-f_{2}(D(G(P))), f_{3}(f_{4}(p))) \right ] \\
            +  \mathbb{E}_{p\sim \mathcal{P},l\sim \mathcal{I}}\left [ f_{1}(D(G(P))-f_{2}(D(l)), f_{5}(f_{4}(p))) \right ] 
    \end{split}
\end{equation}
Where $f_{1}$ is mean square error loss. $f_{2}$ performs mean operation. $f_{4}$ patches the input photo with the same size as outputs of discriminator $D$, $f_{3}$ and $f_{5}$ are used to fill tensor with scalar value 1 and 0, respectively. 

\textbf{Pixel-wise Loss.} Since we use paired samples to train our model, we can introduce the $L_{1}$ loss \cite{chen2018sketchygan} (compared to $L_{2}$ loss, it introduce less blurriness \cite{yi2017dualgan}) to measure the error between the generated and ground truth character line drawings, which can make translated drawings follow the domain $I$ distribution. The function is defined as,  

\begin{equation}
    \mathcal{L}_{pixel-wise} = \mathbb{E}_{p,l}\left [ \left \| G(p) - l \right \|_{1}  \right ] 
\end{equation}

The final objective loss function of our P2LDGAN is given by,
\begin{equation}
    \mathcal{L}_{P2LDGAN} = \lambda_{1}\mathcal{L}_{G}^{P2LDGAN} + \lambda_{2}\mathcal{L}_{D}^{P2LDGAN} + \lambda_{3}\mathcal{L}_{pixelwise}
\end{equation}

In our experiments, we set the weights $\lambda_{1} = 1$, $\lambda_{2} = 0.5$, and $\lambda_{3} = 100$, respectively.

\section{Experiments}

In this section, we first describe the implementation details and evaluation metrics. Then we evaluate the performance of our P2LDGAN by making quantitative and qualitative comparisons against state-of-the-art models as well as reporting human perceptual study. Finally, we conduct ablation study to verify the effectiveness of designed component of our proposed approach.

\subsection{Implementation Details}
We implement the photo-to-line drawing translator using PyTorch. All experiments are conducted on a single NVIDIA GeForce RTX 3090 GPU. For both the generator and discriminator network, we use the Adam optimizer with $\beta_{1}=0.5$ and $\beta_{2}=0.999$. The initial learning rate is set to 0.0002. We train the model for 200 epochs with a mini batch size of 1. All input images and ground truth are aligned and resized to 512 $\times$ 512 pixels.

\subsection{Evaluation Metrics}
To measure the quality of synthesized line drawing images, three commonly-used similarity metrics, namely Fr$\acute{e}$chet Inception Distance (FID) \cite{heusel2017gans} and Structural Similarity Metric (SSIM) \cite{wang2004image} and Peak Signal-to-noise Ratio (PSNR) \cite{chen2020puppeteergan}, are adopted to quantitatively evaluate the performance of previous methods and our proposed P2LDGAN. Where the FID calculates the distribution distance between the set of line drawing images created from input photos and the corresponding ground truth drawings (the smaller the FID value is, the better the drawing quality will be ), while SSIM describes the similarity between the generated image and ground truth (Higher SSIM value indicates better result). PSNR computes the intensity difference between predicted and ground truth images (Larger PSNR scores means smaller difference between images) \cite{pang2021image}. 

\subsection{Comparison with State-of-the-Art Methods}
To demonstrate the superior performance of our P2LDGAN model, we make the quantitative and qualitative comparisons with five state-of-the-art networks, including one existing neural style transfer method (i.e. Gatys \cite{gatys2016image}) as well as four general image-to-image translation approaches, CycleGAN \cite{zhu2017unpaired},  DiscoGAN \cite{kim2017learning}, UNIT \cite{liu2017unsupervised}, pix2pix \cite{isola2017image}, MUNIT \cite{huang2018multimodal}. It is worth noting that we also use paired data to train unsupervised methods.

\subsubsection{Qualitative Comparisons}
\begin{figure*}[htbp]
  \centering
  \includegraphics[width=\textwidth]{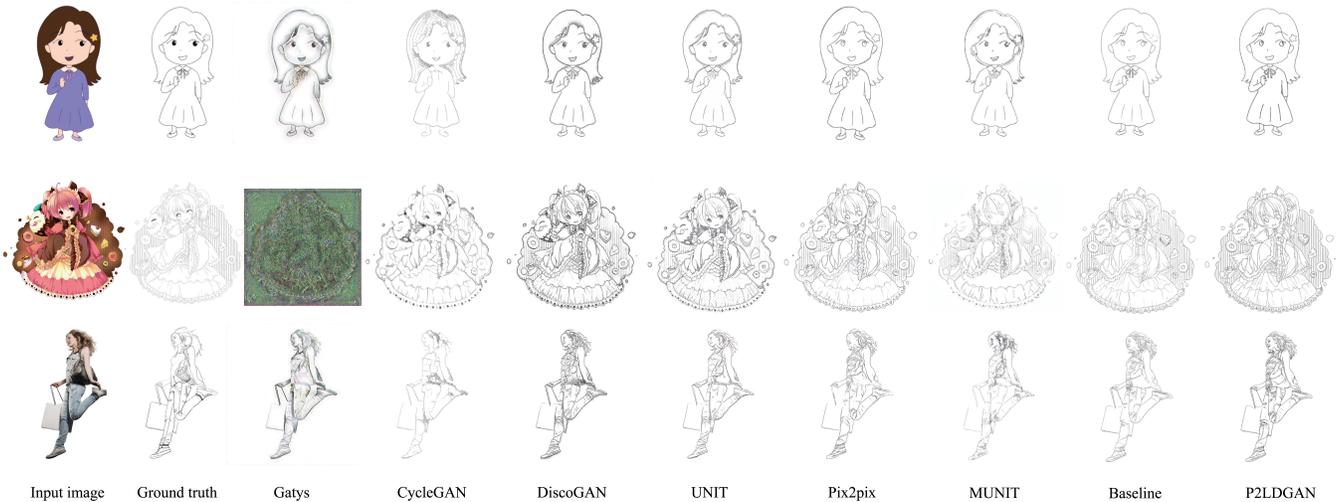}
  \caption{Qualitative examples of different models for character image-to-line drawing generation.}
  \label{fig_experiment}
\end{figure*}

We qualitatively evaluate the performance of our proposed P2LDGAN by 
comparing with state-of-the-art competitors on our testing data. Figure \ref{fig:topbanner} and Figure \ref{fig_experiment} showcase the visual examples. From these results, we can conclude the following findings.  

The character line drawing is essentially composed of abstract lines without any texture information. However, the example-based Gatys \cite{gatys2016image} produces global gray-like feeling  \cite{li2019im2pencil} drawings (shown in the third column) with much texture filled in, which are far apart from the line drawing distribution. It is mainly because of the usage of Gram matrix \cite{yi2020unpaired}. On the contrary, our learning model can generate clear and natural lines with less texture.

For CycleGAN \cite{zhu2017unpaired}, it heavily blurs the details, such as the boy's cloth and shoes in the first row of Figure \ref{fig:topbanner} and introduces great distortions in various areas (e.g., the girl's hair and bag in the third row of Figure \ref{fig_experiment}), which results in visually unappealing line quality.  While pix2pix \cite{isola2017image} succeeds in creating somewhat acceptable perceptual appearance, but it suffers from jagged boundaries, incomplete lines, and details loss. For example,  as can be seen from second row of Figure \ref{fig:topbanner}, pix2pix can not construct the nose for the man. In comparison, our P2LDGAN is able to alleviate such unpleasant artifacts, creating better looking line drawings with more details and structural information preserved.

We also compare our model to DiscoGAN \cite{kim2017learning}, UNIT \cite{liu2017unsupervised} and MUNIT \cite{huang2018multimodal}. Although they can capture the underlying character line styles, and produces more reliable results, the drawings yielded by them are corrupted by messy boundaries and over-smoothed lines. Our approach can give more precise lines to improve visual quality, being coherent with ground truth line drawings. 

In addition, most of these methods do not take full advantage of semantic information, they, therefore, fails to learn the remarkable characteristics of input photo, and generate natural line drawings with less visible artifacts. For example, from the third rows of Figure \ref{fig:topbanner} and Figure \ref{fig_experiment}, it can be explicitly stated that these models could not generate the delicate human face structures. 

In summary, (1) the proposed P2LDGAN significantly outperform state-of-the-art methods in terms of visual quality, details preservation and artifacts reduction. (2) Our method works well on both simple and complex character line drawings learning. 

\subsubsection{Quantitative Comparisons}

To quantify the realism and faithfulness of generated character line drawings \cite{gao2020sketchycoco}, we conduct objective evaluation by computing the average scores of three measurement methods, FID, SSIM and PSNR, on our testing set. The quantitative comparison results with state-of-the-art GAN-based methods are summarized in Table \ref{tab:quantativeresult}. As we can clearly observe, (1) our architecture achieves the best FID, SSIM and PSNR values, significantly outperforming the previous competitors from the realism and faithfulness point of view. (2) Specifically, the lowest FID score suggests that our generated line drawings are closest to ground truth drawings distribution, while the highest SSIM and PSNR scores further show the maximum similarity between our results and ground truths. (3)In summary, the quantitative experiments demonstrate the effectiveness and superiority of our model in synthesizing the high-quality character line drawings, which are consistent with visual results.

\begin{table}[]

\caption{Quantitative comparisons with state-of-the-art models on our introduced dataset. $\downarrow$ indicates lower value is better, while $\uparrow$ means higher score is better.}
\begin{tabular}{cccc} \hline
Methods &  FID $\downarrow$   & SSIM $\uparrow$ & PSNR $\uparrow$\\ \hline
Gatys \cite{gatys2016image}  & 164.1 & 0.6295 & 16.411\\
CycleGAN \cite{zhu2017unpaired}  & 61.8 & 0.8335 & 21.262\\
DiscoGAN \cite{kim2017learning}  & 54.5 & 0.8664 & 20.664\\
UNIT \cite{liu2017unsupervised} & 52.5 & 0.8478 & 20.712\\
pix2pix \cite{isola2017image}  & 52.2 & 0.8917 & 22.711\\
MUNIT \cite{huang2018multimodal}  & 57.1 & 0.8461 & 20.995\\
\hline
Our baseline &50.5 & 0.8984 & \textbf{23.372}\\ 
P2LDGAN & \textbf{47.4} & \textbf{0.9020} & 22.929\\
\hline
\end{tabular}

\label{tab:quantativeresult}
\end{table}

\subsubsection{Human Perceptual Study}
Actually, character image/photo-line drawing translation is a highly subjective task, we, therefore, perform human perceptual studies to perceptually evaluate the ability of our model to generate better-looking drawings.

\textbf{Participant.} 30 participants jointed our user studies, including 20 students without any artistic knowledge, and 10 professional students with at least three years of drawing experience. 

\textbf{Data.} This experiment covers 50 pairs of well-aligned character photos and line drawings we randomly select from the testing set. We use our proposed models and the state-of-the-art networks described in previous section to translate the photos into line drawings. And the ground truth drawings are used as reference images.  

\textbf{Setup.} We show an input photo, ground truth and eight generated line drawings to participants, who are asked to rank the similarity between ground truth and resulting drawings with 4 meaning highest rank, and 1 indicating lowest one \cite{yi2020unpaired}. The users are also asked to rate the overall quality (whether there are noise, deformation, unclear or unstructured lines, incomplete areas, as well as other artifacts in the drawings)  \cite{wang2020learning} on a scale of 1 to 5 scores (Higher score denotes better quality). 

\textbf{Result and Discussion.} We finally collected 1,500 votes in total. The percentage of line drawings being ranked 4 and the average scores of drawing quality are computed and summarized in  Table \ref{tab:userstudy}. From these results, we can report that (1)our method receives the highest values in both evaluation criteria. (2) It performs favorably over these state-of-the-arts frameworks in terms of user preferences. (2) To conclude, our method can be able to learn structural line representation, generating high-quality character line drawings.  

\begin{table*}[]

\caption{Results of human perceptual study. The first row gives the percentage of line drawings being as Rank 4 for each method. The second row presents the average value of overall quality. }
\begin{tabular}{ccccccccc} \hline
Methods &  Gatys \cite{gatys2016image}  & CycleGAN \cite{zhu2017unpaired}  &DiscoGAN \cite{kim2017learning}  & UNIT \cite{liu2017unsupervised} &  pix2pix \cite{isola2017image}  & MUNIT \cite{huang2018multimodal}  & Our baseline & P2LDGAN\\ \hline
Similarity (Rank 4) &  8.2\% & 6.7\% &63.9\% & 35.3\% & 59.5\% & 43.7\% & 67.7\% & \textbf{68.0\%}\\
Overall quality & 2.796 & 2.490 & 4.435 & 3.391 & 4.162 & 3.837 & 4.4723 & \textbf{4.530}\\
\hline
\end{tabular}

\label{tab:userstudy}
\end{table*}

\section{Ablation Study and Discussion}
\begin{figure}[htbp]
  \centering
  \includegraphics[width=0.5\textwidth]{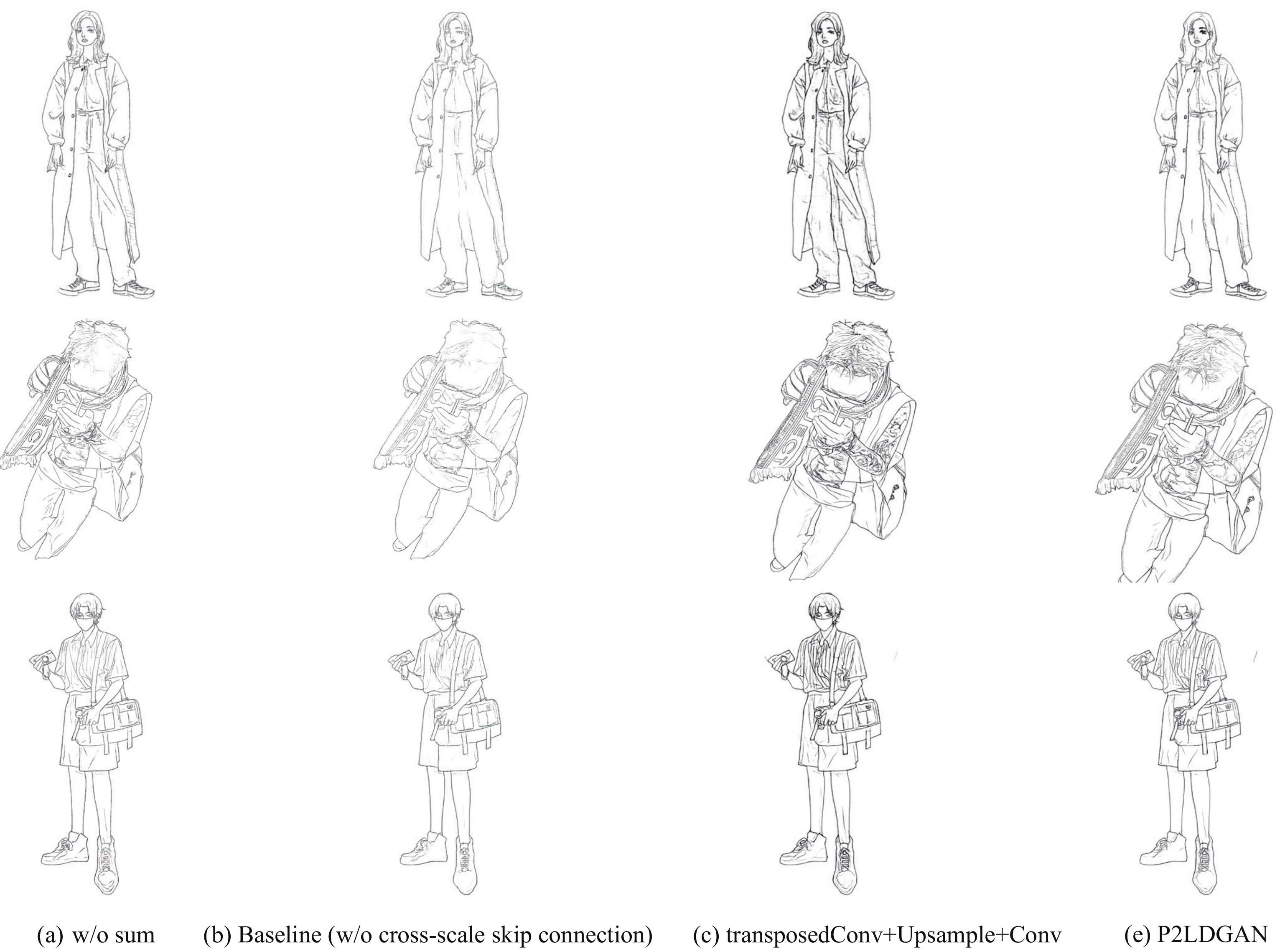}
  \caption{Qualitative results of ablation study.}
  \label{fig_ablation}
\end{figure}
\begin{table}[]

\small
\caption{ Ablation study results.}
\begin{tabular}{ccccc} \hline
Methods &  Parameters & FID $\downarrow$   & SSIM $\uparrow$ & PSNR $\uparrow$\\ \hline

Baseline & 67.2M & 50.5& 0.8984 & 23.372\\
w/o sum & 86.0M & 48.1 & 0.9008&23.322\\
DeConv + Upsample + Conv &103.2M & 47.4 & 0.8994 & 22.222\\
P2LDGAN & 81.1M & 47.4 & 0.9020 & 22.929 \\
\hline
\end{tabular}

\label{tab:ablationeresult}
\end{table}
We conduct the following ablation experiments to study the contributions of the key factors.

\subsection{Analysis of fusion strategy}
We replace the sum operation in cross-scale skip connections modules with concatenation. From Figure \ref{fig_ablation}, we can report that the usage of concatenation achieves acceptable results but having unclear boundaries, some roughtness and noiseness. The numerical values in Table \ref{tab:ablationeresult} also demonstrate the influence of fusion strategies had on drawings quality.

\subsection{Analysis of generator}
In order to validate the effectiveness of our cross-scale skip connections model in keeping fine details,  We perform quantitative and qualitative comparisons between our P2LDGAN with/without using cross-scale skip connections model (our baseline). From the visual results, it can be reported that without our connection model, the generated line drawings have many structural details loss (e.g., hair area in the second row), resulting in poor visual effects, while our P2LDGAN could restore these delicate structures, producing robust and better looking results.

We also replace our decoding layer with a DeConvolution + Upsample + Convolution block. As can be found in Figure \ref{fig_ablation} and Table \ref{tab:ablationeresult}, the network using replaced decoder produces the character line drawings that are similar to or even better than our proposed models. However, it requires more parameters and computation consumption than P2LDGAN.

Summarily, our well-designed P2LDGAN shows superior performance in translating character images/photos into high-quality line drawings with better detailed structure, clear lines, and fewer artifacts.

\section{Conclusion, limitation and future work}
\begin{figure}[htbp]
  \centering
  \includegraphics[width=0.5\textwidth]{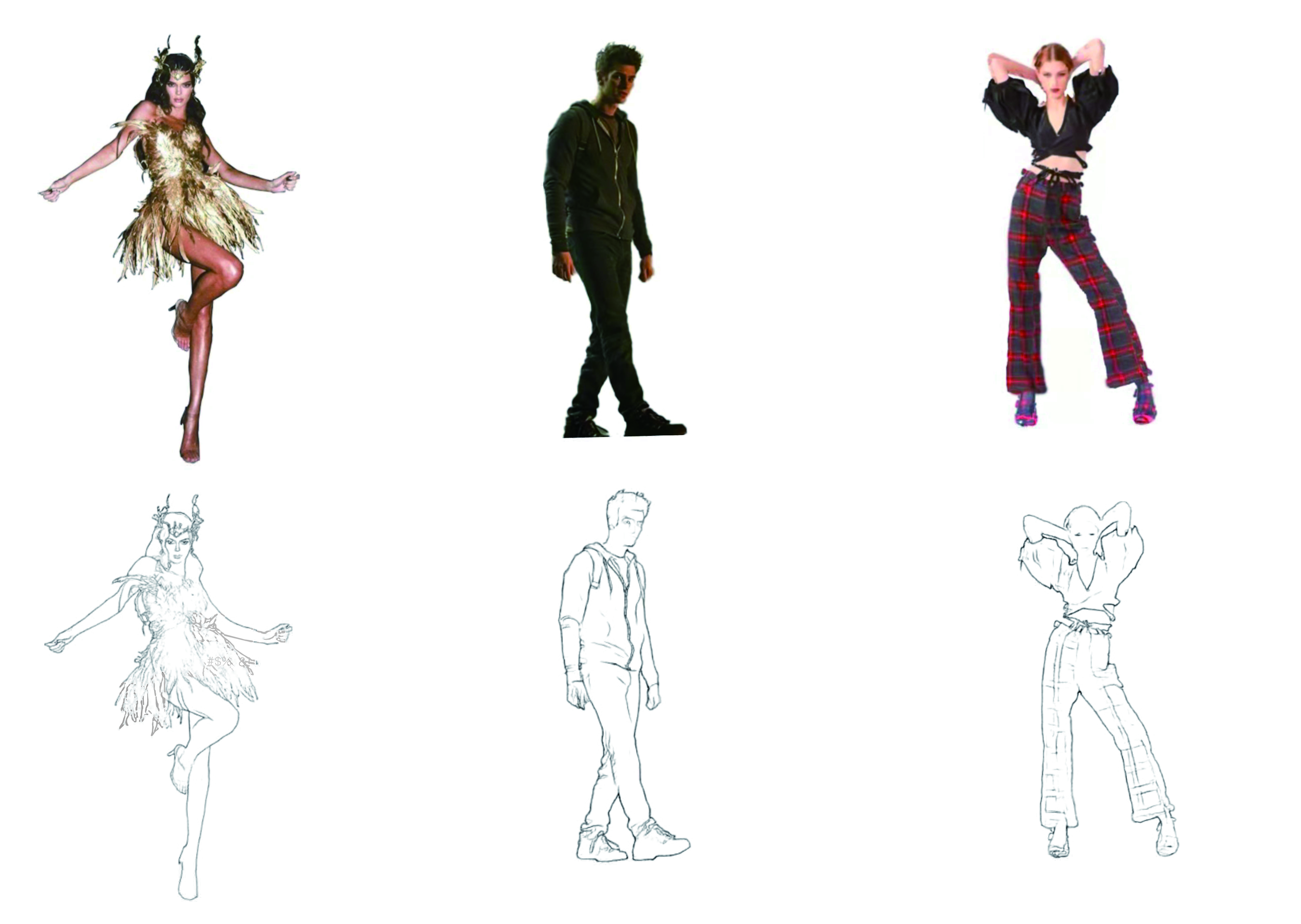}
  \caption{Failure cases.}
  \label{fig_failure}
\end{figure}
In this paper, GANs based image-to-image translation tasks provide a solution to our character line drawing generation problem. We present a novel joint-geometric-semantic driven GAN model, named P2LDGAN, to discover a mapping function that learns the relations between a character image/photo and its translated line drawing representation. We construct a dataset consists of 1,532 pairs of character images/photos and hand-drawn line drawings for training and validating our solution. 
Quantitative and qualitative comparisons with state-of-the-art methods, human evaluation as well as ablation studies show the superiority of our P2LDGAN framework in generation quality. 

\textbf{Limitation.} Figure \ref{fig_failure} presents some failure examples. It can be observed that (1) when the photos are overexposed or underexposed, some areas that are too bright (e.g., the cloth in the first column) or too dark (e.g., the facial area in the second column) would be missing in the generated line drawings. (2) The low-quality original photos, such as these with blurry texture, would cause noisy and messy lines, failing to preserve fine details, like the third column. 
These problems may be considered in our future study, and we will also investigate how to improve the realism and faithfulness of the real human line drawings. 

\bibliographystyle{ACM-Reference-Format}
\bibliography{sample-base}

\end{document}